\documentclass[usenatbib]{mn2e}

\usepackage[british]{babel}
\usepackage{txfonts}
\let\umu=\muup
\usepackage{bm}
\usepackage{graphicx}

\begin{document}

\title[Imaging the cold molecular gas in SDSS J$1148+5251$ at $z=6.4$]{Imaging the cold molecular gas in SDSS J$\bm{1148+5251}$ at $\bm{z=6.4}$}
\author[Stefan et al.]{Irina I. Stefan,$^1$\thanks{E-mail: iis21@cam.ac.uk} Chris L. Carilli,$^{2,1}$ Jeff Wagg,$^{3,1}$ Fabian Walter,$^4$ Dominik A. Riechers,$^5$ 
\newauthor Frank Bertoldi,$^6$ David A. Green,$^1$ Xiaohui Fan,$^7$ Karl Menten,$^8$ Ran Wang$^{9,2}$ \\
$^1$Cavendish Laboratory, 19 J. J. Thomson Ave., Cambridge, CB3 0HE, UK \\
$^2$National Radio Astronomy Observatory, Socorro, NM, USA \\
$^3$Square Kilometre Array Organization, Jodrell Bank Observatory, Lower Withington, Macclesfield, Cheshire, SK11 9DL, UK \\
$^4$Max-Planck Institut f\"{u}r Astronomie, K\"{o}nigstuhl 17, D-69117 Heidelberg, Germany\\
$^5$Department of Astronomy, Cornell University, 220 Space Sciences Building, Ithaca, NY 14853, USA \\
$^6$Argelander-Institut f\"{u}r Astronomie, Universit\"{a}t Bonn, Auf dem H\"{u}gel 71, Bonn, D-53121, Germany \\
$^7$Steward Observatory, University of Arizona, 933 North Cherry Avenue, Tucson, AZ 85721, USA \\
$^8$Max-Planck-Institut f\"{u}r Radioastronomie, Auf dem H\"{u}gel 69, 53121 Bonn, Germany \\
$^9$Kavli Institute for Astronomy and Astrophysics, Peking University, Beijing 100871, China}

\maketitle

\begin{abstract}
We present Karl G. Jansky Very Large Array (VLA) observations of the CO ($J = 2 \rightarrow 1$) line emission towards the $z = 6.419$ quasar SDSS J$114816.64+525150.3$ (J$1148+5251$). The molecular gas is found to be marginally resolved with a major axis of $0\farcs9$ (consistent with previous size measurements of the CO ($J = 7 \rightarrow 6$) emission). We observe tentative evidence for extended line emission towards the south west on a scale of $\sim 1\farcs4$, but this is only detected at $3.3\sigma$ significance and should be confirmed.  The position of the molecular emission region is in excellent agreement with previous detections of low frequency radio continuum emission as well as \mbox{[C\,{\sc ii}]} line and thermal dust continuum emission. These CO ($J = 2 \rightarrow 1$) observations provide an anchor for the low excitation part of the molecular line SED. We find no evidence for extended low excitation component, neither in the spectral line energy distribution nor the image. We fit a single kinetic gas temperature model of 50~K. We revisit the gas and dynamical masses in light of this new detection of a low order transition of CO, and confirm previous findings that there is no extended reservoir of cold molecular gas in J$1148+5251$, and that the source departs substantially from the low $z$ relationship between black hole mass and bulge mass. Hence, the characteristics of  J$1148+5251$ at $z  = 6.419$ are very similar to $z \sim 2$ quasars, in the lack of a diffuse cold gas reservoir and kpc-size compactness of the star forming region. 
\end{abstract}


\section{Introduction}
Studies of molecular lines towards high redshift galaxies provide powerful diagnostics on the properties of the dense interstellar medium (ISM) in the early Universe. As cool gas is the fuel for star formation, such studies are probes of galaxy formation and evolution and help us to understand the origins of today's galaxies (for a recent review, see \citealp{2013ARA&A..51..105C}).

At $z=6.419$, SDSS J$114816.64+525150.3$ (hereafter J$1148+5251$) is the most distant quasar discovered in the Sloan Digital Sky Survey (SDSS, \citealp{2003AJ....125.1649F}), and remains among the most distant known. It is a radio quiet quasar \citep{2004AJ....128..997C, 2003AJ....126...15P} with an estimated supermassive black hole mass of $3 \times 10^9~\rmn{M}_{\sun}$ accreting at the Eddington limit, as found from \mbox{Mg\,{\sc ii}} emission lines \citep{2003ApJ...587L..15W}. Optical spectra of the source show a saturated Gunn--Peterson trough due to a partially neutral intergalactic medium, providing some of the first evidence from the earliest epochs of galaxy formation and cosmic reionization at $z > 6$ \citep{2005AJ....129.2102W}. J$1148+5251$ was the first $z > 6$ quasar to be detected as a hyperluminous infrared galaxy (HyLIRG) (infrared luminosity $> 10^{13}$ L$_{\sun}$, \citealp{2003A&A...406L..55B}) through millimetre bolometer observations. It also remains the most distant detection of molecular gas to date, with multiple detections of CO rotational transition lines from $J = 3 \rightarrow 2$ up to $J = 7 \rightarrow 6$ \citep{2003Natur.424..406W,2003A&A...409L..47B, 2009ApJ...703.1338R} and the possible detection of the CO ($J = 17 \rightarrow 16$) transition \citep{2014MNRAS.445.2848G}. These revealed a massive molecular gas reservoir of $\ga 10^{10}~M_{\sun}$ \citep{2003A&A...409L..47B, 2003Natur.424..406W, 2007ApJ...666L...9C} centred on the active galactic nucleus (AGN) and extended on scales of 5 kpc \citep{2004ApJ...615L..17W, 2009ApJ...703.1338R}.

\mbox{[C\,{\sc ii}]} observations trace a bright $1.5$ kpc atomic gas region with active star formation apparently taking place at an estimated star-formation rate (SFR) surface density of $\sim 1000~M_{\sun}$ yr$^{-1}$ kpc$^{-2}$ \citep{2009Natur.457..699W}. Based on \mbox{[C\,{\sc ii}]}, \citet{2015A&A...574A..14C} suggest the presence of a high velocity complex outflow extending to projected radii of $\sim 30$ kpc. J$1148+5251$ has a far-infrared (FIR) luminosity of $2.7 \times 10^{13}~L_{\sun}$ \citep{2013ApJ...772..103L}, corresponding to thermal emission from $4.2 \times 10^8~M_{\sun}$ of warm dust \citep{2006ApJ...642..694B}.

J$1148+5251$ has served as the archetype for extreme starburst-AGN at high redshifts, representing the early, coeval formation of supermassive black holes and their stellar host galaxies during the epoch of reionization.

In this paper, we present observations with the Karl G. Jansky Very Large Array (VLA) of the CO ($J = 2 \rightarrow 1$) line emission in J1148+5251. These observations have a number of significant advantages over the previous VLA observations. Low order CO transitions are critical to constrain the total molecular gas mass in galaxies, and in particular, to look for massive reservoirs of cold gas fuelling star formation \citep{2006ApJ...650..604R, 2011ApJ...739L..32R, 2011ApJ...733L..11R, 2011MNRAS.412.1913I}. The lowest previously detected transition from J$1148+5251$ was CO ($J = 3 \rightarrow 2$) \citep{2003Natur.424..406W}, with rather shallow limits on the CO ($J = 1 \rightarrow 0$) line emission \citep{2003A&A...409L..47B}. CO ($J = 3 \rightarrow 2$) can be substantially sub-thermally excited, even in starburst galaxies, with a typical brightness temperature ratio of 0.66 for submillimetre galaxies (SMGs), and 0.6 for high $z$ main sequence galaxies \citep{2013ARA&A..51..105C}. Conversely, the CO ($J = 2 \rightarrow 1$) to CO ($J = 1 \rightarrow 0$) ratio is almost always thermal ({i.e.} the brightness temperature is constant due to thermalized emission), even in high-$z$ main sequence galaxies, with ratios of 0.85 in SMGs and 0.9 even in high $z$ main sequence galaxies. Hence, CO ($J = 2 \rightarrow 1$) provides an excellent route to study cool gas reservoirs, being four times stronger (in erg s$^{-1}$ Hz$^{-1}$) than CO ($J = 1 \rightarrow 0$), but still tracing the colder gas \citep{2014arXiv1409.8158D}.

We use the capabilities of the VLA to image the full emission line from J$1148+5251$. The previous CO ($J = 3 \rightarrow 2$) observations used the old VLA correlator, which had limited bandwidth and channelization, leading to an observation that missed the line wings beyond $\pm 150$ km s$^{-1}$, and had only 12 channels across the centre of the line.

And lastly, we obtain sensitive observations of the continuum emission at a rest frame frequency of 230 GHz, corresponding to an emission regime that would be dominated by the cold dust.

We adopt a standard $\Lambda$CDM cosmological model with $H_{0} = 67.3$ km s$^{-1}$ Mpc$^{-1}$, $\Omega_{\Lambda} = 0.685$, $\Omega_{\rmn{M}} = 0.315$ \citep{2014A&A...571A..16P}. At a redshift of $z = 6.419$, this implies an age of the universe of 851 Myr and a scale of $5.623$ kpc arcsec$^{-1}$.

\section{Observations}\label{Observations}
We use the VLA in C and D configurations to observe the CO ($J = 2 \rightarrow 1$) ($\nu_\rmn{rest} = 230.538$ GHz) emission line towards J1148$+$5251. The line is redshifted to $\nu = 31.074$ GHz at $z=6.419$. Observations in the D configuration were performed between November 2011 and January 2012 for a total of $\sim 15$ hours on source. Observations in the C configuration followed between January and April 2012 for a total of $\sim 9$ hours on source. Since these observations were taken as JVLA early science during the initial commissioning of the new correlator, we chose to observe topocentric frequencies. Furthermore, the observations span a six month range that involved major changes to the correlator between C and D configurations. Hence, we did not attempt to doppler track, but have summed the topocentric spectra. This leads to at most a $20$ km s$^{-1}$ smearing of spectral features, which is comparable to a single channel in our final analysis. This smearing will have no effect on our subsequent analysis.

The standard calibrator 3C286 was used as a primary flux and bandpass calibrator and the nearby source J1153$+$4931 was observed approximately every five and a half minutes for pointing, secondary amplitude and phase calibration.

These observations were taken during the commissioning phase of the VLA correlator, and hence correlator capabilities were evolving rapidly. The D array observations were obtained when full channelization (64 channels) was available, but only for a 128 MHz bandwidth. This bandwidth is easily adequate to cover any reasonable line emission ($\pm 600$ km s$^{-1}$), but does not allow for a sensitive continuum measurement. The later C array observations had the same line coverage, but also had close to 1 GHz of bandwidth off-line, providing a sensitive continuum observation.

The data were calibrated, imaged and analysed using standard techniques in NRAO's CASA package\footnote{The Common Astronomy Software Applications package, http://casa.nrao.edu} \citep{2007ASPC..376..127M}. Some difficulties were encountered due to the presence of two bright sources near the half power position of the primary beam. Small inherent pointing inaccuracies resulted in these sources appearing as time-variant leading to significant residual sidelobes even after deconvolution. To solve this issue we performed continuum subtraction in the $uv$-plane using the outer most channels (4 to 9 and 57 to 62) of the 128 MHz band centred on the emission line (channels 23 to 43). 

We imaged the C-array data separately using a close-to-`natural' weighting scheme characterized by a `robust' parameter \citep{Briggs1995} of 1. We then repeated the process for the D-array data only. We also created a combined data set from the C and D-array data and imaged it with different weightings to explore spatial resolution versus signal-to-noise optimization. A close-to-`uniform' weighting scheme (robust $=-1$ in CASA) was used to achieve higher spatial resolution, a close-to-`natural' robust $=1$ was used to obtain higher sensitivity, and a good trade-off between resolution and sensitivity was achieved with a middle-ground robust $=0$. This way we produced maps with a large range in resolution, as shown in Table \ref{robust}.

\begin{table}
\caption{Noise and resolution of a number of maps obtained with the VLA data presented in this paper.}
\label{robust}
\begin{tabular}{c c c c} 
\hline
Arrays & `robust' & rms per channel & Restoring \\
used & parameter & (19.309 km s$^{-1}$) & beam \\
\hline
C & 1 & 95 $\umu$Jy beam$^{-1}$ & $0\farcs76 \times 0\farcs69$ \\
D & 1 & 68 $\umu$Jy beam$^{-1}$ & $2\farcs53 \times 2\farcs28$ \\
C\&D & $-1$ & 126 $\umu$Jy beam$^{-1}$ & $0\farcs55 \times 0\farcs52$ \\
C\&D & 0 & 73 $\umu$Jy beam$^{-1}$ & $0\farcs82 \times 0\farcs75$ \\
C\&D & 1 & 56 $\umu$Jy beam$^{-1}$ & $1\farcs75 \times 1\farcs62$ \\
\hline
\end{tabular}
\end{table}

\section{Results}\label{Results}
\begin{figure}
	\centerline{\includegraphics[width=\columnwidth,trim=1.8cm 17.2cm 4cm 1.3cm, clip=true]{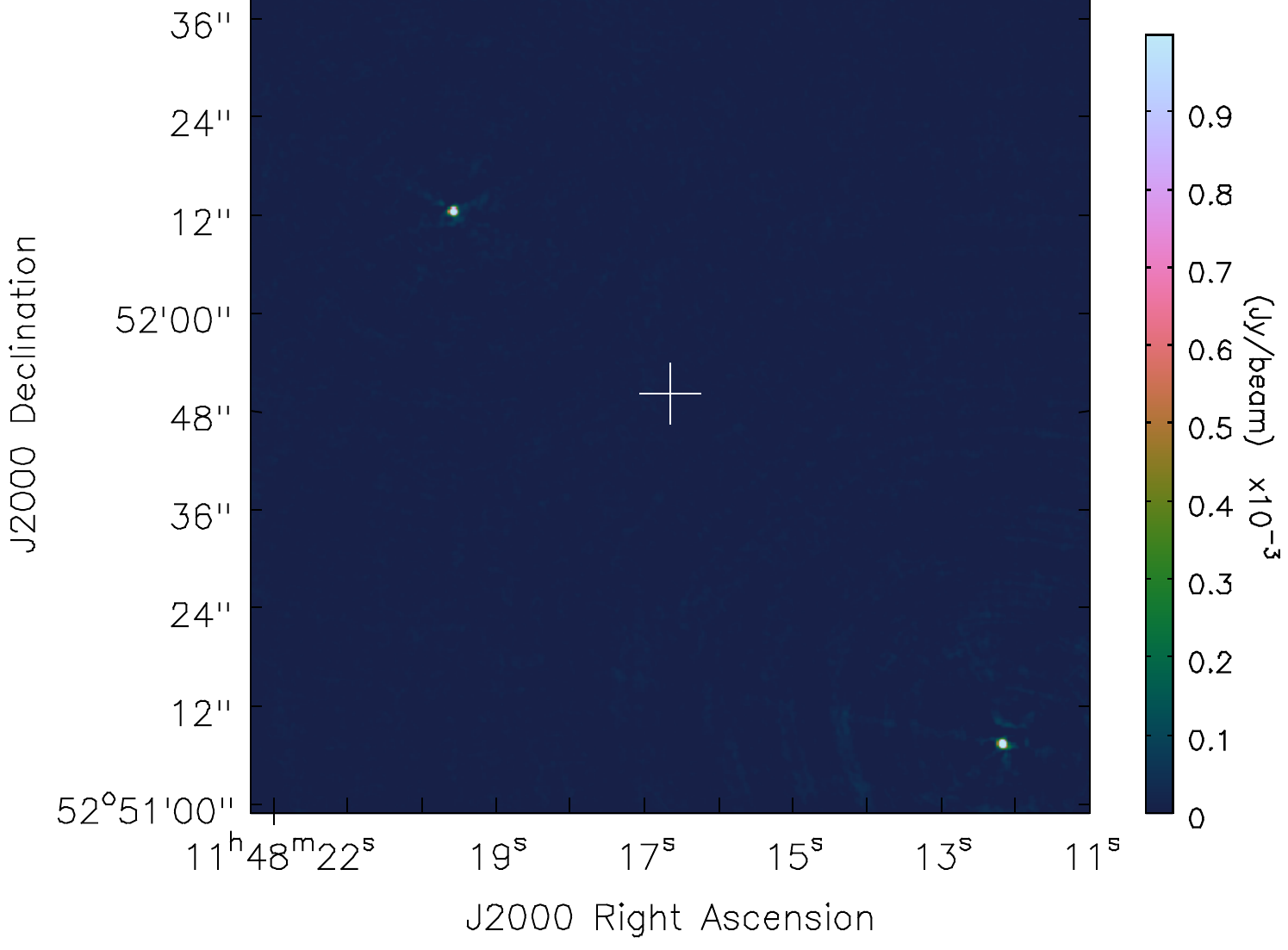}}
	\caption{Continuum map of the C-array data. The rms is $6.4~\umu$Jy. Two lower redshift sources to the north west and the south east of J$1148+5251$ dominate the emission all over the map. The cubehelix colour scheme \citep{2011BASI...39..289G} was used.}
	\label{Cfull_cont}
\end{figure}

The field covered by these observations also includes two bright radio sources \citep{1995ApJ...450..559B}, one $\sim35\arcsec$ to the NE of J$1148+5251$ and one $\sim40\arcsec$ to the south west, both at much lower redshift \citep{1995ApJ...450..559B}. They have been mentioned in previous studies of J1148+5251 \citep[e.g.][]{2003A&A...406L..55B}. Figure \ref{Cfull_cont} is a continuum map of the C-array data only, showing these sources. We measure a non-primary beam corrected flux density of $2.79 \pm 0.05$ mJy for the north eastern source and of $2.98 \pm 0.06$ mJy for the south western. We perform $uv$-continuum subtraction to remove the effects of these sources from subsequent analysis.

\begin{figure}
	\centerline{\includegraphics[width=\columnwidth,trim=0.5cm 0cm 1cm 1cm, clip=true]{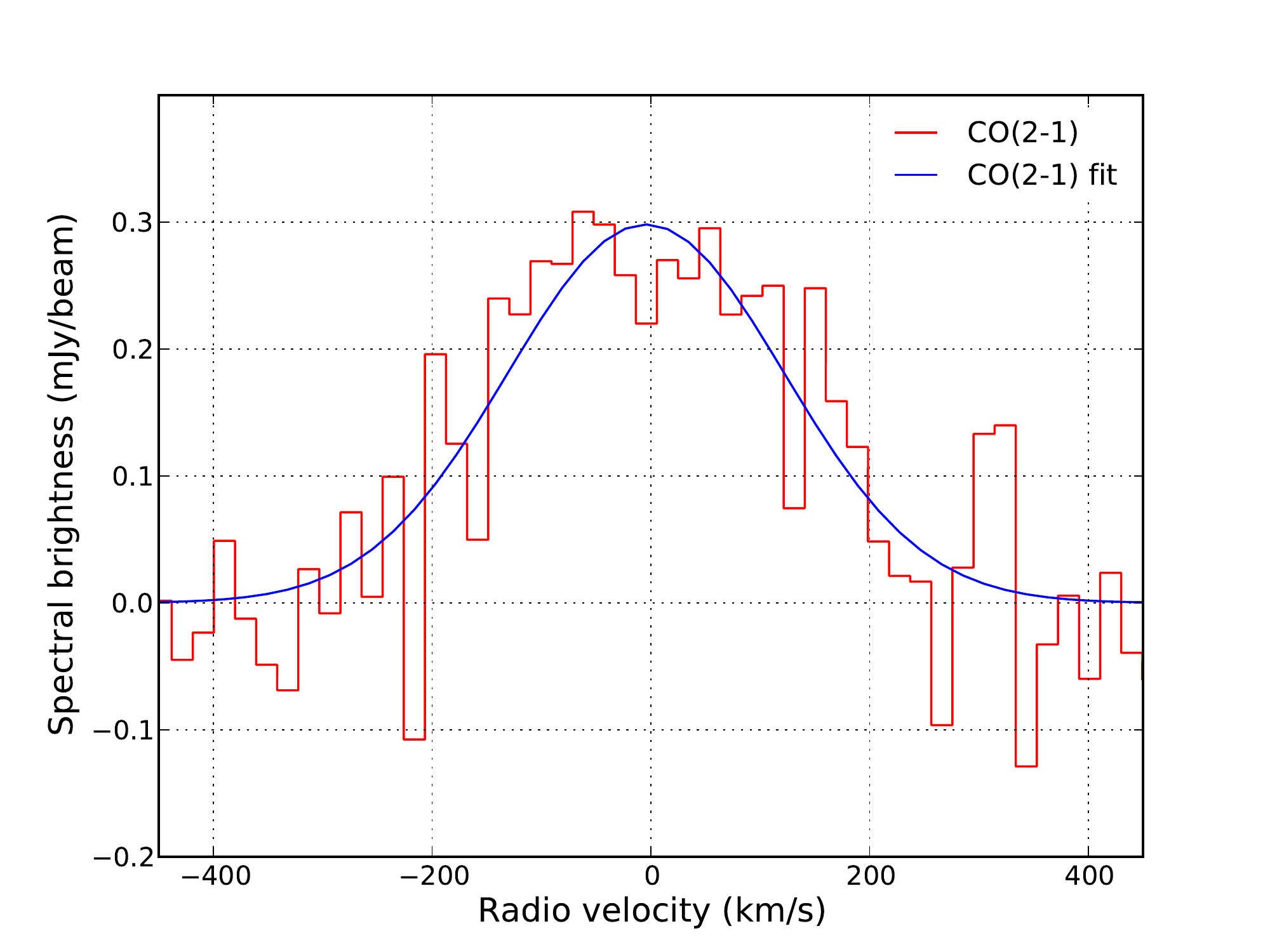}}
	\caption{Spectrum of the CO ($J = 2 \rightarrow 1$) line emission towards J$1148+5251$ obtained with VLA D array observations at a resolution of 19.335 km~s$^{-1}$ ($2$ MHz). The blue line shows a Gaussian fit to the data. The velocity scale is relative to the redshifted CO ($J = 2 \rightarrow 1$) frequency at $z = 6.419$.}
	\label{spectrum_fit}
\end{figure}

We clearly detect the CO ($J = 2 \rightarrow 1$) emission towards J$1148+5251$. Figure \ref{spectrum_fit}, shows the CO ($J = 2 \rightarrow 1$) emission spectrum obtained from the D-array data. Fitting a Gaussian to the emission gives a peak intensity of $I_{\nu} = 0.298 \pm 0.022$ mJy beam$^{-1}$ and a FWHM of $298 \pm 26$ km s$^{-1}$ centred on $-5 \pm 15$ km s$^{-1}$. This implies an integrated line flux of $94.6 \pm 7.7$ mJy beam$^{-1}$ km s$^{-1}$ and a line luminosity of $L'_{CO} = 3.2 \pm 0.3 \times 10^{10}$ K km s$^{-1}$ pc$^{2}$.

Constraining the source size in CO ($J = 2 \rightarrow 1$) emission is difficult because of the moderate signal-to-noise and non-ideal $uv$ coverage, both due to a shorter-than-requested observation time on the full C-array. The combined C\&D observations yield a synthesized beam with broad wings, depending on weighting. To further explore the issue of the CO ($J = 2 \rightarrow 1$) emission size, we carried out two tests.

\begin{figure}
	\centerline{\includegraphics[width=\columnwidth,trim=0.5cm 0cm 1cm 1cm, clip=true]{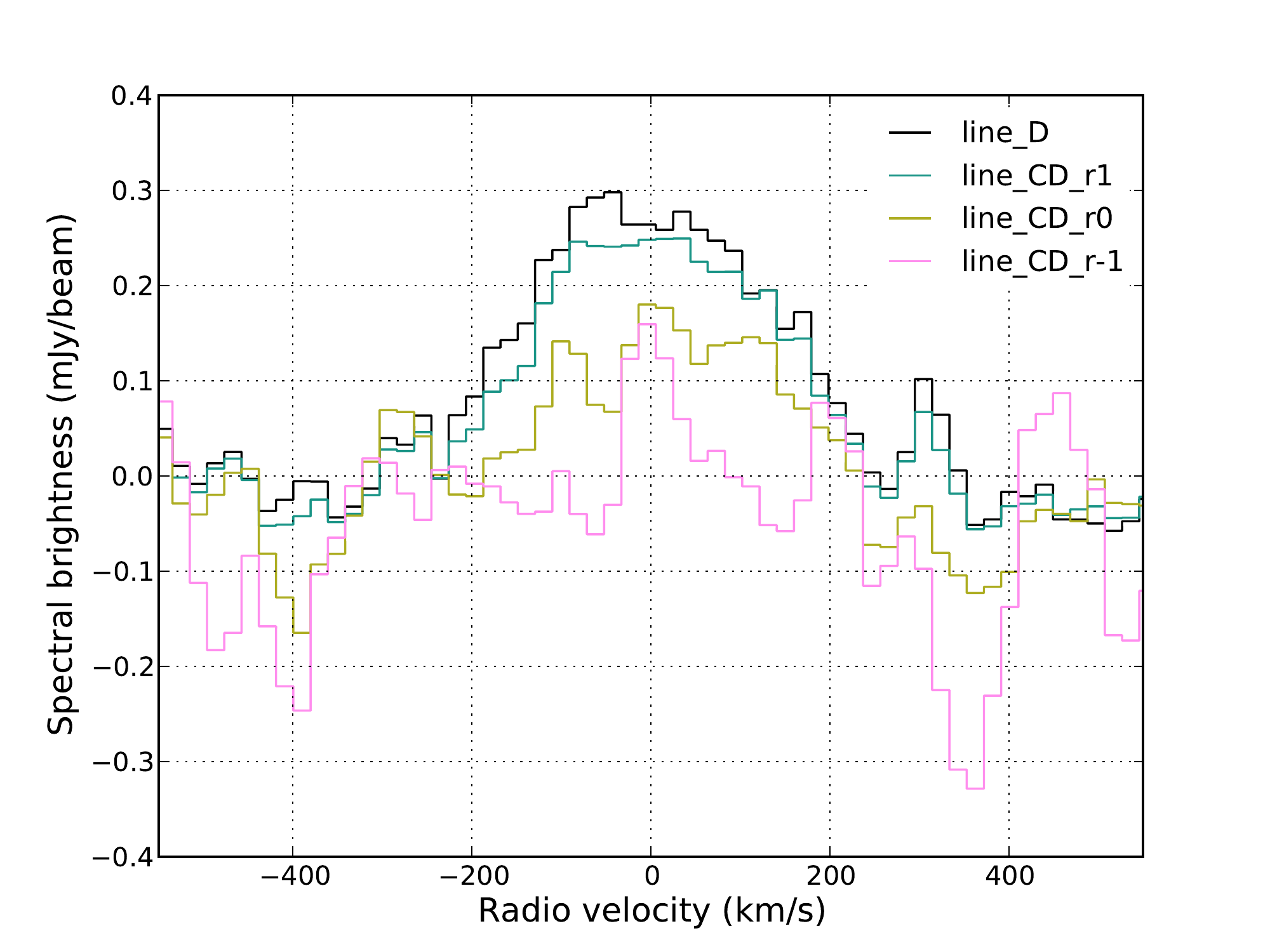}}
	\caption{Spectra of the CO ($J = 2 \rightarrow 1$) emission towards J$1148+5251$ from: D-array only data imaged with a weighting parameter robust of 1 (black), C\&D-array data imaged with robust of 1 (blue), C\&D-array data imaged with robust of 0 (yellow) and C\&D-array data imaged with robust of $-1$ (pink).}
	\label{4spectrum}
\end{figure}

First, we looked at point-source line profiles at various resolutions. We considered the position of the peak emission from the D-array only map (derived from a Gaussian fit) as the position of the point source and took line profiles from the D-array only and C\&D-array data imaged with three different weightings (robust $=1$,~$0$ and $-1$). These spectra are shown in Fig. \ref{4spectrum}. Moving from a close-to-`natural' (robust $=1$) weighting towards a close-to-`uniform' (robust $=-1$) one there is a clear decrease in the peak emission, as it drops from $\sim 0.30$ to $\sim 0.15$ mJy beam$^{-1}$. Fitting Gaussians to each of the spectra, we find a difference of $\sim 60\%$ between the D-array only spectral amplitude and the C\&D-array robust $=-1$ spectral amplitude. The restoring beams for each of these maps are listed in Table \ref{robust}. Figure \ref{4spectrum} shows that the peak flux decreases with increasing resolutions, and hence the source is resolved at higher resolution.

\begin{figure}
	\centerline{\includegraphics[width=\columnwidth,trim=2cm 17.3cm 3.7cm 1cm, clip=true]{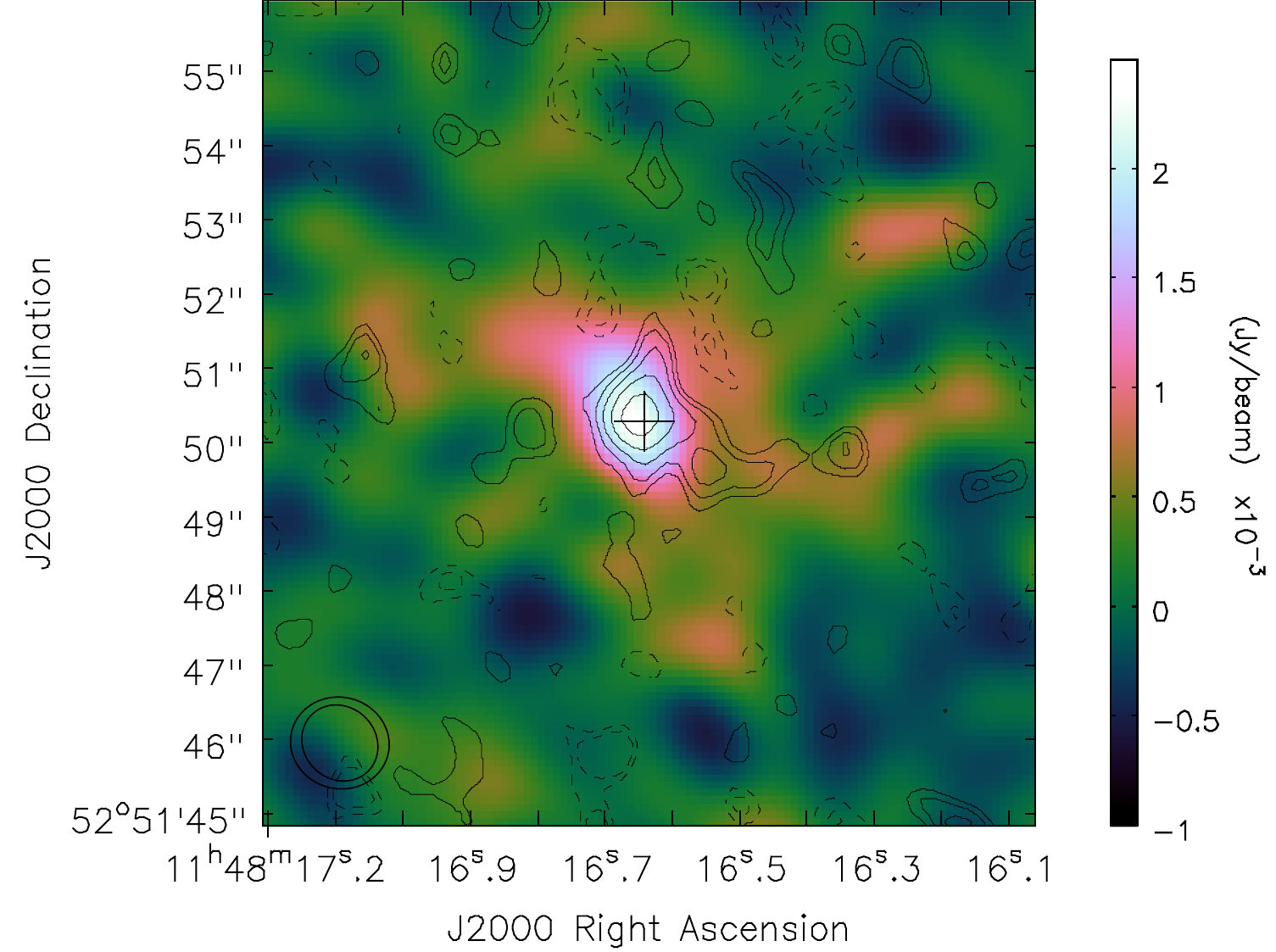}}
	\caption{Velocity-integrated VLA (C \& D configurations) contour map of the CO ($J = 2 \rightarrow 1$) line emission towards J$1148+5251$ overlaid on to \mbox{[C\,{\sc ii}]} emission observed by \citet{2015A&A...574A..14C}. The \mbox{[C\,{\sc ii}]} emission is integrated over a larger velocity range that the CO one. The cross marks the position of the optical quasar \citep{2005AJ....129.2102W}. Contours are at ($-2,-1.41, 1.41, 2, 2.82, 4, 5.64, 8$)$\times\sigma$ ($1\sigma = 25 \umu$Jy beam$^{-1}$). The synthesized beam of the CO data is $1\farcs07 \times 0\farcs97$ and that of the \mbox{[C\,{\sc ii}]} data is $1\farcs33 \times 1\farcs23$.}
	\label{total_Cii}
\end{figure}

Second, we imaged the combined C\&D-array dataset. This is problematic as well due to the different weightings required for optimal signal-to-noise by data from the two arrays and the broad line wings resulting from having significantly more time on-source with the D array versus the C array. To avoid these issues, we CLEANed the integrated line emission (velocities between $\sim -216$ km s$^{-1}$ and $\sim 189$ km s$^{-1}$) to a depth where the synthesized beam's broad wings are removed and explored different Briggs weights. Figure \ref{total_Cii} shows contours of the J1148+5251 velocity-integrated CO ($J = 2 \rightarrow 1$) emission imaged with a robust of $0.25$ overlaid on to the \mbox{[C\,{\sc ii}]} emission observed by \citet{2015A&A...574A..14C}. The spatial resolution of the CO ($J = 2 \rightarrow 1$) observations is $1\farcs07 \times 0\farcs97$, just slightly higher than that of the \mbox{[C\,{\sc ii}]} observation, as shown in the bottom left corner of the map. 

The main emission region is less than $1$ arcsec in extent, although there is some diffuse emission to the south west of the source at $3 \sigma$ which needs further confirmation. A formal elliptical Gaussian fit to the robust $= 0.5$ image returns a deconvolved major axis of $0\farcs89 \pm 0\farcs46$, and an undetermined minor axis and position angle. So we marginally resolve the emission in one direction and measure a size consistent with that found by \cite{2009ApJ...703.1338R} for the CO ($J = 7 \rightarrow 6$) emission.

The peak flux of $157 \pm 17~\umu$Jy beam$^{-1}$ is measured close to the position of the optical quasar (see cross in Fig. \ref{total_Cii}; \citealp{2005AJ....129.2102W}) and agrees very well with the position of the \mbox{[C\,{\sc ii}] emission}. 

Much deeper observations are needed to determine the source morphology as a function of velocity.

We do not detect any continuum in the CO ($J = 2 \rightarrow 1$) observations down to a limit of $1\sigma = 6.7~\umu$Jy.

\section{Analysis}\label{Analysis}
The line width of $298.05 \pm 25.87$ km s$^{-1}$ we derive for the CO ($J = 2 \rightarrow 1$) emission agrees well with the width measured in previous studies \citep{2003A&A...409L..47B, 2009ApJ...703.1338R, 2009Natur.457..699W}. Hence, the low and high order transitions have similar line profiles in J$1148+5251$.

The FIR-to-radio spectral energy distribution fit from \citet{2008ApJ...687..848W} uses an emissivity index $\beta = 1.6$, a dust temperature $T = 55$~K \citep{2006ApJ...642..694B} and a FIR-to-radio ratio $q = 2.34$ \citep{2001ApJ...554..803Y} and predicts an observed flux density of $\sim 9~\umu$Jy, with a large uncertainty depending on $q$. Our non-detection is just below the predicted value. This result confirms a $\beta$ value that is at least as steep as $1.6$, and possibly steeper, thereby strengthening the argument for little cold dust in J$1148+5251$.

\subsection{Gas mass estimate}
Far infrared CO luminosity ratios and linewidths in quasars at high redshifts are similar to what is observed in local ultra-luminous infrared galaxies (ULIRGs) \citep{2005ARA&A..43..677S, 2013ARA&A..51..105C} so we adopt the CO luminosity to H$_{2}$ mass conversion factor of $\alpha = 0.8~\rmn{M}_{\sun}$~(K km s$^{-1}$ pc$^2$)$^{-1}$ used for ULIRGs \citep{1998ApJ...507..615D}. From \cite{2013ARA&A..51..105C}, the line ratio $L'_\rmn{CO{2-1}}/L'_\rmn{CO{1-0}}$ is typically very close to unity for high redshift galaxies, so we can approximate $L'_\rmn{CO{1-0}}$ as $3.2 \pm 0.3 \times 10^{10}$ K km s$^{-1}$ pc$^{2}$. Thus, from the CO (J$ = 2 \rightarrow 1)$ luminosity we derive a total molecular gas mass of $M_{\rmn{H}_{2}} = 2.6 \times 10^{10}~\rmn{M}_{\sun}$.

\subsection{Dynamical mass estimate}
Assuming a rotating disc geometry for the line-emitting gas in J$1148+5251$ we can estimate the dynamical mass of the emitting region using $M_\rmn{dyn} \approx 1.16 \times 10^{5} v_\rmn{circ}^{2}~D~\rmn{M}_{\sun}$, where $D$ is the diameter in kpc we measured for the emission and $v_\rmn{circ}$ is the FWHM in km s$^{-1}$ of the line profile. Following \citet{2013ApJ...773...44W} we use the approximation $v_\rmn{circ}~\sin$~$i = 0.75$~FWHM$_{\rmn{CO} (J = 2 \rightarrow 1)}$. Taking the size of the source as 8.1 kpc (the major axis) implies $M_\rmn{dyn}~\sin^{2}i = 5.3 \times 10^{10}~\rmn{M}_{\sun}$. For an inclination angle of $63\degr$ as found from the ratio of the minor to major axis, the dynamical mass is $M_\rmn{dyn} = 6.7 \times 10^{10}~\rmn{M}_{\sun}$, in line with previously derived values. \citet{2004ApJ...615L..17W} quote $M_\rmn{dyn} = 4.5 \times 10^{10}~\rmn{M}_{\sun}$ (or $M_\rmn{dyn} = 5.5 \times 10^{10}~\rmn{M}_{\sun}$ if accounting for an inclination of $65\degr$) within a disc of diameter 5 kpc. \citet{2009ApJ...703.1338R} use the \mbox{[C\,{\sc ii}]} line emission to find a dynamical mass of $M_\rmn{dyn} = 1.5 \times 10^{10}~\rmn{M}_{\sun}$ for a region with radius $1.5$ kpc. It has to be noted that there is substantial uncertainty to modelling assumptions (rotating disc), and model parameters (size, inclination) in our dynamic mass estimate. High resolution imaging in \mbox{[C\,{\sc ii}]} or CO at much higher signal-to-noise is required to better understand the galaxy dynamics in J$1148+5251$.

\subsection{CO Line Excitation}
We have expanded the large velocity gradient (LVG) model of the line excitation in J$1148+5251$ calculated by \citet{2009ApJ...703.1338R} in which the kinetic gas temperature and density are considered as free parameters. A fixed $3:1$ H$_{2}$ ortho-to-para ratio was used due to the relative statistical weights of the symmetrical and antisymmetrical eigenstates of the wavefunction. This model uses the \citet{2001Flower} CO collision rates and a cosmic microwave background temperature of $20.25$ K at $z=6.42$. We kept the parameters adopted by \citet{2009ApJ...703.1338R}: a CO abundance per velocity gradient of [CO]$/(\rmn{d}v/\rmn{d}r) = 1 \times 10^{-5}$ pc/(km s$^{-1}$) \citep{2005A&A...438..533W, 2006ApJ...650..604R, 2007A&A...467..955W} and a CO disc radius of 2.5 kpc \citep{2009ApJ...703.1338R, 2004ApJ...615L..17W}.

\begin{figure}
	\centerline{\includegraphics[width=\columnwidth]{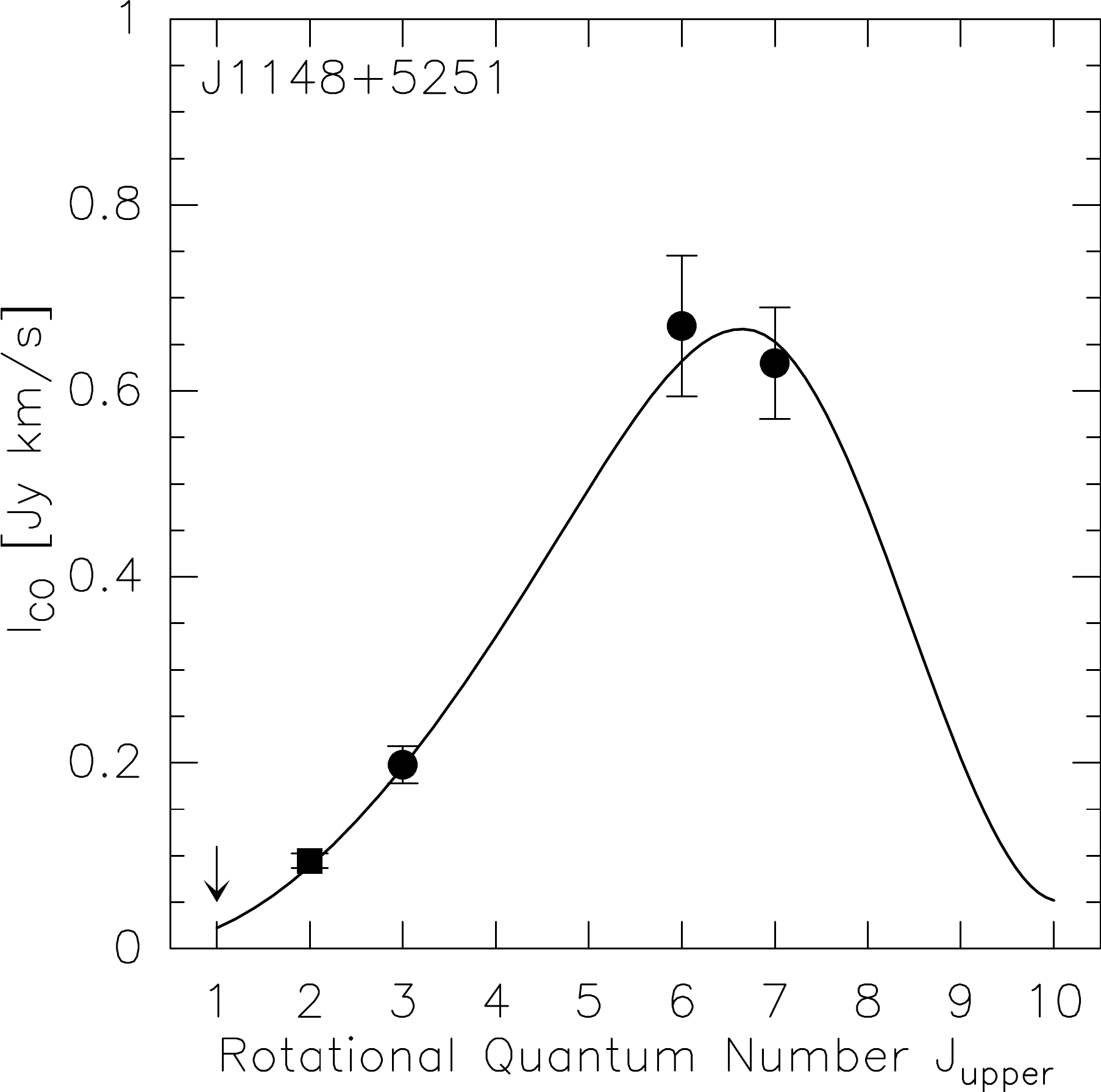}}
	\caption{CO excitation ladder (symbols) and LVG model (line) for J$1148+5251$. The upper limit for CO ($J = 1 \rightarrow 0$) and the (continuum corrected) data point for CO (J$ = 6 \rightarrow 5$) are taken from \citet{2003A&A...406L..55B}. The CO ($J = 7 \rightarrow 6$) (continuum corrected) data point is taken from \citet{2009ApJ...703.1338R} an the CO (J$ = 3 \rightarrow 2$) data point (corrected for missing flux in the line wings) is taken from \citet{2003Natur.424..406W}. The square marks the data point from the current observations. The model predicts kinetic temperatures and gas densities of $T_\rmn{kin} = 50$ K and $\rho_\rmn{gas}$(H$_{2}$)$ = 10^{4.2}$ cm$^{-3}$.}
	\label{COladder}
\end{figure}

The CO ($J = 2 \rightarrow 1$) data point fits well on the curve described by the best solution (Fig. \ref{COladder}). This was obtained for a spherical, single-component model with a CO disc filling factor of 0.16, $T_\rmn{kin} = 50$ K, and $\rho_\rmn{gas}$(H$_{2}$)$ = 10^{4.2}$ cm$^{-3}$.

\section{Discussion}\label{Discussion}

We have observed the CO ($J = 2 \rightarrow 1$) transition towards J$1148+5251$ and find supporting evidence that J$1148+5251$ is similar to lower redshift quasars in terms of its high CO line excitation and compact size. We marginally resolve the main emission region to have a major axis of $0\farcs89 $. We also detect tentative evidence for diffuse emission extending to the south west.

Considering the line excitation in J$1148+5251$, our data are used to provide further constraints on previous LVG models, resulting in average physical conditions that are similar to those observed in other high-redshift quasars \citep{2007ASPC..375...25W, 2009ApJ...703.1338R}. We find $T_\rmn{kin} = 50$~K extending from low to high order transitions. As such, it is possible that the emission in the low order transitions of CO may be associated with the same warm, highly excited star-forming gas traced by higher-\textit{J} line transitions. This would support previous claims that the mid-\textit{J} CO lines (\textit{J}$\leq 3$) may also be good indicators of the total amount of cool molecular gas in high redshift quasars \citep{2006ApJ...650..604R, 2011ApJ...739L..32R}. 

In the local Universe, the traditional view of ultraluminous infared galaxies (ULIRGs) and AGN has been that these objects represent different phases in an evolutionary sequence \citep{1988ApJ...325...74S}. A heavily dust obscured galaxy undergoing vigorous star-formation will be observed as a ULIRG, while a buried AGN continues to grow a supermassive blackhole. This may eventually give rise to a far-infrared luminous quasar, followed by an optically luminous quasar with only a modest gas reservoir remaining. This scenario may be similar  to what is observed in high-redshift quasars, like  J$1148+5251$, which may have gone through an earlier phase of starburst activity before the AGN had formed. Extended reservoirs of cold molecular gas have been detected via CO line emission in $z \simeq 2$ SMGs \citep{2010ApJ...714.1407C, 2010MNRAS.404..198I, 2011MNRAS.412.1913I, 2010ApJ...720L.131R, 2011ApJ...733L..11R, 2011ApJ...739L..31R}, suggesting that they may be in the process of forming through gas-rich mergers. A population of $z \sim 6$ SMGs has recently been discovered  (e.g. Riechers et al.\ 2013; Weiss et al.\ 2013), so it is possible that these may represent an early phase of the luminous $z\sim 6$ quasars such as J$1148+5251$. These objects are likely to evolve into the most massive early-type galaxies observed in the local Universe. However, in the case of the single known $z \sim 6$ SMG, the lack of a hot dust component detected in the Herschel bands makes this object inconsistent with such a scenario. More work is required to establish the validity of the $z\sim$6 SMGs and $z\sim6$ quasars connection.

We calculate new estimates of the gas and dynamical mass from the CO ($J = 2 \rightarrow 1$) line luminosity. Our results agree with previous measurements, but are less prone to biases due to high order molecular line excitation. The gas mass we estimate is $\sim 63\%$ of the dynamical mass, suggesting a low value for $\alpha$. This supports a departure from the $M_\rmn{BH} - \sigma_\rmn{bulge}$ relationship, in that the $M_\rmn{dyn}$ of the host galaxy is well below that expected from the low $z$ relation.

\section{Acknowledgements}
IIS thanks the Science \& Technology Facilities Council for a studentship. The National Radio Astronomy Observatory is a facility of the National Science Foundation operated under cooperative agreement by Associated Universities, Inc. We thank Claudia Cicone and Roberto Maiolino for helpful discussions and for providing the \mbox{[C\,{\sc ii}]} image.

\end{document}